\documentclass[a4 paper, 10pt]{article}
\usepackage[lmargin=1in,rmargin=1in,tmargin=1in,headsep=.2in]{geometry}
\usepackage{graphicx}
\usepackage{forloop}
\usepackage{etoolbox}
\usepackage{calc}
\usepackage{wrapfig,lipsum,booktabs} 
\usepackage{xtab} 
\usepackage[dvipsnames]{xcolor}
\usepackage[normalem]{ulem}
\usepackage{sectsty}
\usepackage{hyperref}
\hypersetup{
    colorlinks=true,
    urlcolor=blue,
    linkcolor=black,
    citecolor = black}

\usepackage{amsmath,amsfonts,amssymb,amsthm,epsfig,epstopdf,url,array}
 \usepackage{makeidx,epsfig,lscape}
 \usepackage{xcolor,pict2e}
\usepackage{upgreek}
\usepackage{float}

\theoremstyle{definition}

\begin{document}
\vspace*{0cm}
{\noindent\huge\bf Stochastic Simulation of Emission Spectra and Classical Photon Statistics of Quantum Dot Superluminescent Diodes}\\[0.5cm]
{\bf\large Kai Niklas Hansmann and Reinhold Walser}\\[0.25cm]
Technische Universit\"{a}t Darmstadt, Institut f\"{u}r Angewandte Physik, Hochschulstra\ss e 4a, 64289 Darmstadt, Germany\\
Email: kai.hansmann@physik.tu-darmstadt.de\\[0.5cm]
We present a stochastic procedure to investigate the correlation spectra of quantum dot superluminescent diodes. The classical electric field of a diode is formed by a polychromatic superposition of many independent stochastic oscillators. Assuming fields with individual carrier frequencies, Lorentzian linewidths and amplitudes we can form any relevant experimental spectrum using a least square fit. This is illustrated for Gaussian and Lorentzian spectra, Voigt profiles and box shapes. Eventually, the procedure is applied to an experimental spectrum of a quantum dot superluminescent diode \cite{Hartmann2017} which determines the first- and second-order temporal correlation functions of the emission. We find good agreement with the experimental data and a quantized treatment \cite{Friedrich2020}. Thus, a stochastic field represents broadband light emitted by quantum dot superluminescent diodes.
\vspace{0.5cm}\\
\vspace{-0.5cm}
\section{Introduction}

Modern day optical applications like optical coherence tomography \cite{Huang1991,Judson2009} and ghost imaging \cite{Hartmann2017,Pittman1995,Gatti2006} make use of the unique emission properties of spectrally broadband light-emitting quantum dot superluminescent diodes (QDSLD). By using specialized waveguide geometries and gain materials, QDSLDs are able to combine high output intensities, spatially directed emission and spectral widths in the THz regime. Hence, they fill the gap in the family of semiconductor-based optical emitters between coherent laser diodes and incoherent light emitting diodes. After being proposed in 1973 \cite{Lee1973}, research on the characteristics of QDSLDs has been intensified in recent years after Boitier et al. \cite{Boitier2009} enabled the direct measurement of coherence times in the femtosecond regime using two-photon absorption in semiconductors. This research includes experimental studies of emission and photon statistical characteristics of QDSLDs \cite{Jechow2013,Kiethe2017,Blazek2011}, as well as theoretical investigations based on rate equations \cite{Uskov2004,Bardella2009,Forrest2020}, travelling wave approaches \cite{Rossetti2011}, finite element methods \cite{Li2010} and a quantized treatment \cite{Friedrich2020, Hartmann2015}. To this day, effort is being put into developing more efficient and high-powered QDSLDs \cite{Ozaki2019,Forrest2019,Aho2019}.

Adding a new perspective to the investigation of QDSLDs, we discuss a stochastic model for the emission in this article. Stochastic approaches have long proven to have a wide-ranging field of applications in \mbox{biology \cite{Brueckner2019,Fang2019,Vagne2018}}, \mbox{engineering \cite{Nesti2018}}, finance \cite{Kanazawa2018}, quantum many-body physics \cite{Hartmann2019}, soft-matter physics \cite{Qi2020,Liebchen2019}, optics \cite{Walser1994, McIntyre1993} and many more scientific areas. We develop a model to describe experimental emission spectra from a superposition of stochastic fields. Using least square fits, we determine Lorentzian linewidths, carrier frequencies and amplitudes of the individual fields to model experimental spectra. This is illustrated for Gaussian-, Lorentzian-, Voigt- as well as bandpass spectra and applied to the experimental spectrum of a QDSLD \cite{Hartmann2017}. Using numerical simulations, we determine first- and second-order temporal correlation functions of the electric field and calculate the spectral power density of the resulting emission.

The article is organized as follows: the stochastic model of emission spectra is developed in Sec. \ref{sec:model}. It consists of individual classical fields, which are described by a distinct stochastic differential equation. After investigating properties of these fields, relevant spectra are modelled as a superposition. This is applied to a specific experimental spectrum produced by a QDSLD \cite{Hartmann2017}. The model is subsequently used to calculate the emission spectrum of the diode in Sec. \ref{sec:spectrum} and the normalized stationary second-order temporal correlation function in Sec. \ref{sec:secOrdTemp}. A conclusion is given in Sec. \ref{sec:conclusion}. An appendix summarizes the convergence properties of the simulation schemes.

\section{Stochastic model of emission spectra} \label{sec:model}

The classical electric field of a diode results from a superposition of stochastic fields. Hence, the electric field outside of the diode reads
\begin{align}
	\varepsilon_{\text{d}}(t)=\sum_{j=1}^N\varepsilon_j(t),\label{eq:diodeField}
\end{align}
with the number of fields $N$ and $\varepsilon_j(t)$ the $j$-th complex field amplitude.

\subsection{Ornstein-Uhlenbeck process}

An individual classical field $\varepsilon(t)\in\mathbb{C}$ is modelled as a complex Ornstein-Uhlenbeck process \cite{Uhlenbeck1930, Gardiner2009}. This is described by the Ito stochastic differential equation
\begin{align}
	\text{d}\varepsilon(t)=(\text{i}\nu_0-\gamma)\,\varepsilon(t)\,\text{d}t+D\,\text{d}W(t),\label{eq:fieldQD}
\end{align}
with the carrier frequency $\nu_0$, the linewidth $\gamma$, the diffusion constant $D=\sqrt{\gamma I}$, the mean intensity of the electric field $I=\lim_{t\rightarrow\infty}\langle\vert\varepsilon(t)\vert^2\rangle$ and the complex Wiener noise increment $\text{d}W(t)\in\mathbb{C}$, whose properties are given by $\langle\text{d}W(t)\rangle=0$ and $\langle\vert\text{d}W(t)\vert^2\rangle=\text{d}t$.

The stationary first-order temporal correlation function reads \cite{Gardiner2009,Glauber1963}
\begin{align}
	G_{\text{s}}^{(1)}(\tau)=\lim_{t\rightarrow\infty}\left\langle\varepsilon^*(t)\varepsilon(t+\tau)\right\rangle=I\text{e}^{-\gamma\vert\tau\vert-\text{i}\nu_0\tau}.\label{eq:statTempCorCGF}
\end{align}
The spectral power density is given by the Fourier transform
\begin{align}
	S(\nu)=\frac{1}{\sqrt{2\pi}}\int_{-\infty}^{\infty}
\text{d}\tau\;G_{\text{s}}^{(1)}(\tau)\text{e}^{\text{i}\nu\tau}
\end{align}
of eq. (\ref{eq:statTempCorCGF}) in accordance to the Wiener-Khintchine theorem \cite{Wiener1930, Khintchine1934}. This yields
\begin{align}
	S(\nu)=&I\sqrt{\frac{2}{\pi}}\frac{\gamma}{(\nu-\nu_0)^2+\gamma^2},& \frac{1}{\sqrt{2\pi}}\int_{-\infty}^{\infty}\text{d}\nu\;S(\nu)=&I.\label{eq:spectralPowerDensityQD}
\end{align}
Furthermore, the stationary normalized second-order temporal correlation function is given by the Siegert relation \cite{Glauber1963,Mandel1995}
\begin{align}
	g_{\text{s}}^{(2)}(\tau)=\lim_{t\rightarrow\infty}\frac{\langle\varepsilon^*(t)\varepsilon^*(t+\tau)\varepsilon(t+\tau)\varepsilon(t)\rangle}{\langle\varepsilon^*(t)\varepsilon(t)\rangle\langle\varepsilon^*(t+\tau)\varepsilon(t+\tau)\rangle}
	=1+\text{e}^{-2\gamma\vert\tau\vert}.\label{eq:secOrdTempAna}
\end{align}

\subsection{Stochastic simulation}

In addition to analytical results we perform numerical simulations of eq. (\ref{eq:fieldQD}). In order to obtain an efficient simulation procedure we separate the rapid oscillating carrier frequency by the transformation $\varepsilon(t)=\eta(t)\text{e}^{-\text{i}\nu_0 t}$ yielding
\begin{align}
		\text{d}\eta(t)=-\gamma\,\eta(t)\,\text{d}t+D\,\text{d}W(t).
\end{align}
As the diffusion constant $D$ is independent of the electric field amplitude $\eta(t)$, the Euler scheme \cite{Kloeden1999} can be used to achieve strong convergence of order $1.0$ (see Appendix \ref{sec:numSimSDE}). Therefore the electric field amplitude can be simulated iteratively
\begin{align}
	\eta(t_{i+1})=\eta(t_i)-\gamma\eta(t_i)\,\Delta t+D\,\Delta W,\label{eq:simField}
\end{align}
with the discrete time step $\Delta t=t_{i+1}-t_i$ and $\Delta W$ a complex Gaussian random process with mean $\langle\Delta W\rangle=0$ and variance $\langle\vert\Delta W\vert^2\rangle=\Delta t$.

The first-order temporal correlation function of eq. (\ref{eq:statTempCorCGF}) is calculated from a sample average over $M$ realizations
	\begin{align}
		G^{(1)}(\tau)=\frac{1}{M}\sum_{m=1}^M\left(\varepsilon^{(m)}(t_s)\right)^*\varepsilon^{(m)}(t_s+\tau)\label{eq:simTempCor}
	\end{align}
long after the transient regime $t_s\gg 1/\gamma_j$. $\varepsilon^{(m)}(t)$ is the $m$-th realization of the electric field. This result can be used to calculate the spectral power density of the emission using a Fourier transformation.

\begin{figure}[h]
	\centering
	\includegraphics[width=0.45\textwidth]{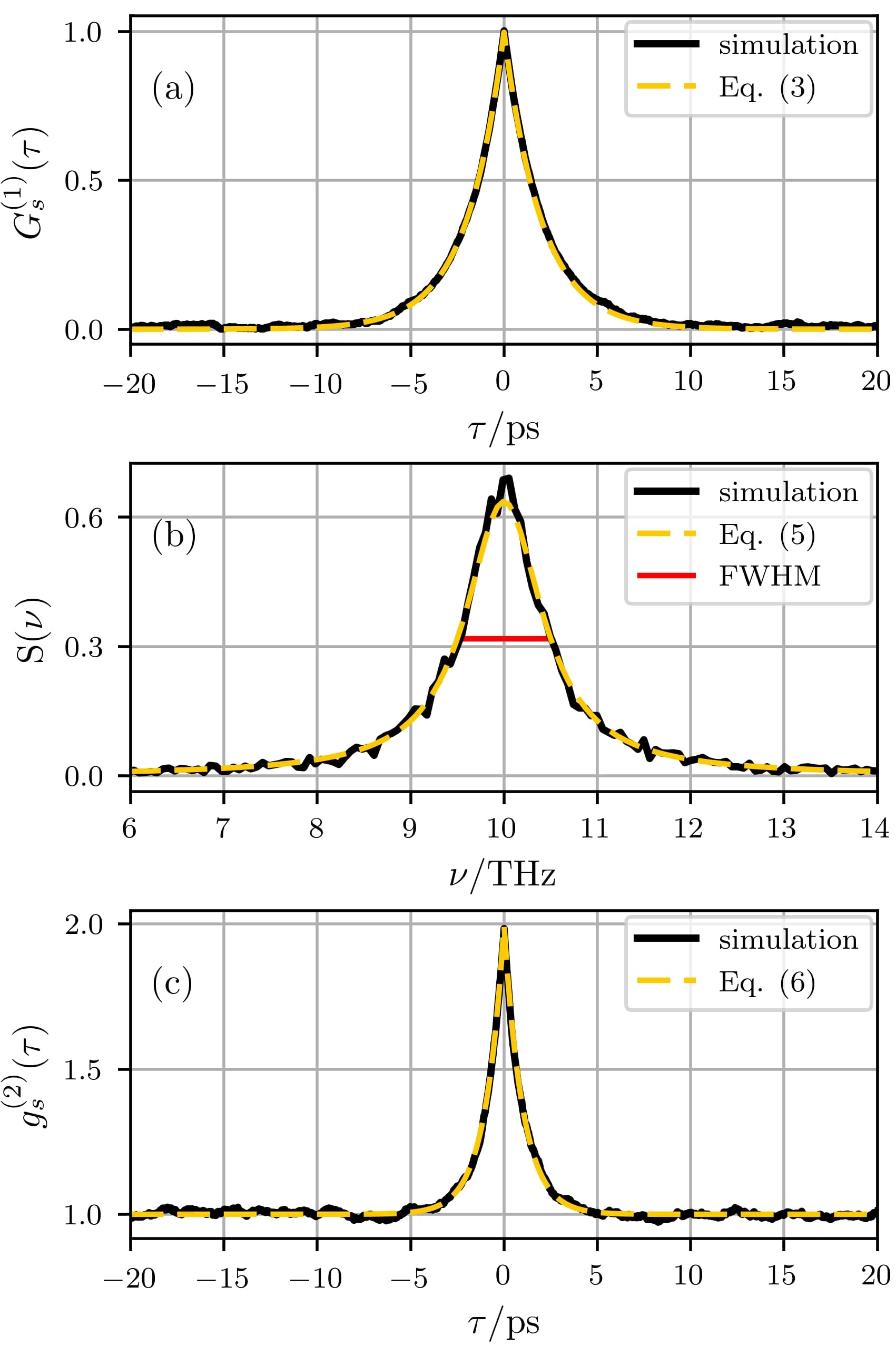}
	\caption{(a) Stationary first-order temporal correlation function $G_{\text{s}}^{(1)}(\tau)$ versus time $\tau$, (b) spectral power density $S(\nu)$ versus \mbox{frequency $\nu$} with width (FWHM) $2\gamma=1$\,THz and (c) stationary normalized second-order temporal correlation function $g_{\text{s}}^{(2)}(\tau)$ versus time $\tau$ for the electric field described by eq. (\ref{eq:fieldQD}). Simulation results (black, solid) and analytical expressions (yellow, dashed) were calculated with $I=1$ and $\nu_0=10$\,THz. The correlation functions were determined using $\Delta t=0.01$\,ps and $M=10^4$ realizations.}
	\label{fig:propertiesSingleQD}
\end{figure}

The determination of the normalized second-order temporal correlation function (\ref{eq:secOrdTempAna}) can be split into two separate calculations. The first-order temporal correlation functions in the denominator can be simulated according to eq. (\ref{eq:simTempCor}), while the second-order correlation function in the numerator can be calculated as
\begin{align}
	G^{(2)}(\tau)=\frac{1}{M}\sum_{m=1}^M\left\vert\varepsilon^{(m)}(t_s+\tau)\,\varepsilon^{(m)}(t_s)\right\vert^2.\label{eq:secOrdTempSim}
\end{align}
Simulation results as well as analytical calculations of the first- and second-order temporal correlation properties of an individual field can be seen in Fig. \ref{fig:propertiesSingleQD}. The simulations show good agreement with the analytical results for the given parameters ($\gamma=0.5$\,THz, $I=1$, $\nu_0=10$\,THz, $\Delta t=0.01$\,ps, $M=10^4$).

\subsection{Modelling of emission shapes}

The emission of a diode (\ref{eq:diodeField}) is described as the superposition of $N$ independent classical fields with individual linewidths $\gamma_j$, mean intensities $I_j$ and central frequencies $\nu_j$. The stationary first-order temporal correlation function reads
\begin{align}
	G_{\text{d}}^{(1)}(\tau)=\lim_{t\rightarrow\infty}\left\langle\varepsilon_{\text{d}}^*(t)\varepsilon_{\text{d}}(t+\tau)\right\rangle=\lim_{t\rightarrow\infty}\sum_{j=1}^{N}\left\langle\varepsilon_j^*(t)\varepsilon_j(t+\tau)\right\rangle.
\end{align}
Thus, the spectral power density is the incoherent sum of the individual spectra
\begin{align}
	S_{\text{d}}(\nu)=\sum_{j=1}^NS_j(\nu).\label{eq:sumLorentizans}
\end{align}

\begin{figure}[H]
	\centering
	\includegraphics[width=0.55\textwidth]{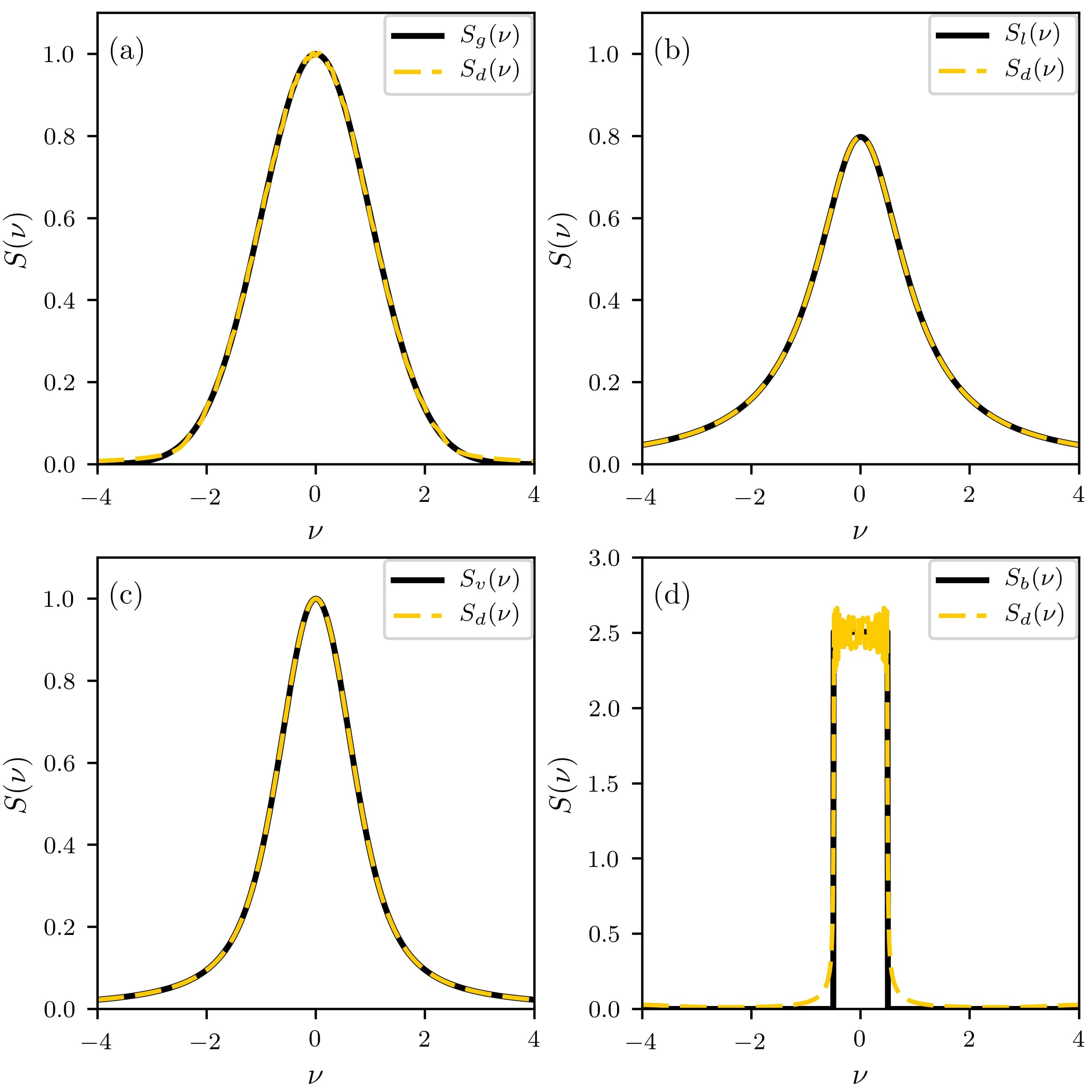}
	\caption{(a) Gaussian spectrum $S_{\text{g}}(\nu)$, (b) Lorentzian spectrum $S_{\text{l}}(\nu)$, (c) Voigt profile $S_{\text{v}}(\nu)$ and (d) bandwidth limited box shape $S_{\text{b}}(\nu)$ versus frequency $\nu$ (black, solid; $\nu_0=0$, $\sigma=1$, $\gamma=1$) modelled according to eq. (\ref{eq:sumLorentizans}) with $N=30$ elementary oscillators (yellow, dashed). The oscillations at the edge of the box are a typical Gibbs phenomenon \cite{Jerri1998}.}
	\label{fig:arbitraryModel}
\end{figure}

This model can be used to approximate a wide range of shapes through the adjustment of the $3N$ free parameters $\gamma_j$, $I_j$ and $\nu_j$ in eq. (\ref{eq:sumLorentizans}) by means of a least square fit, minimizing the error functional
\begin{align}
	e=\sum_{i}(S_{\text{t}}(\nu_i)-S_{\text{d}}(\nu_i))^2\label{eq:error}
\end{align}
for a test spectrum $S_{\text{t}}(\nu)$ at discrete frequencies $\nu_i$. Examples of interest \cite{NIST} are given by Gaussian spectra
	\begin{align}
		S_{\text{g}}(\nu)=\frac{1}{\sqrt{\sigma^2}}\text{e}^{-\frac{(\nu-\nu_0)^2}{2\sigma^2}},
	\end{align}
Lorentzian spectra
	\begin{align}
		S_{\text{l}}(\nu)=\sqrt{\frac{2}{\pi}}\frac{\gamma}{(\nu-\nu_0)^2+\gamma^2},
	\end{align}
Voigt profiles
	\begin{align}
		S_{\text{v}}(\nu)=&\frac{1}{\sqrt{\sigma^2}}\text{Re}\left\{\text{e}^{-z^2}\text{erfc}(-\text{i}z)\right\}, & z=\frac{\nu+\text{i}\gamma}{\sigma\sqrt{2}},
	\end{align}
with the complementary error function $\text{erfc}(z)$, and bandwidth limited box shapes
	\begin{align}
		S_{\text{b}}(\nu)=\left\{\begin{matrix}
				\sqrt{2\pi}/\gamma, & \text{for } \vert\nu\vert\leq\gamma/2,\\
				0, & \text{else.}
		\end{matrix}\right.
	\end{align}
This is illustrated in Fig. \ref{fig:arbitraryModel}. Please note that all spectra are normalized to
\begin{align}
	\frac{1}{\sqrt{2\pi}}\int_{-\infty}^{\infty}\text{d}\nu\;S(\nu)=1.
\end{align}

\subsection{Model of quantum dot superluminescent diode emission} \label{model}

\begin{figure}[H]
	\centering
	\includegraphics[width=0.52\textwidth]{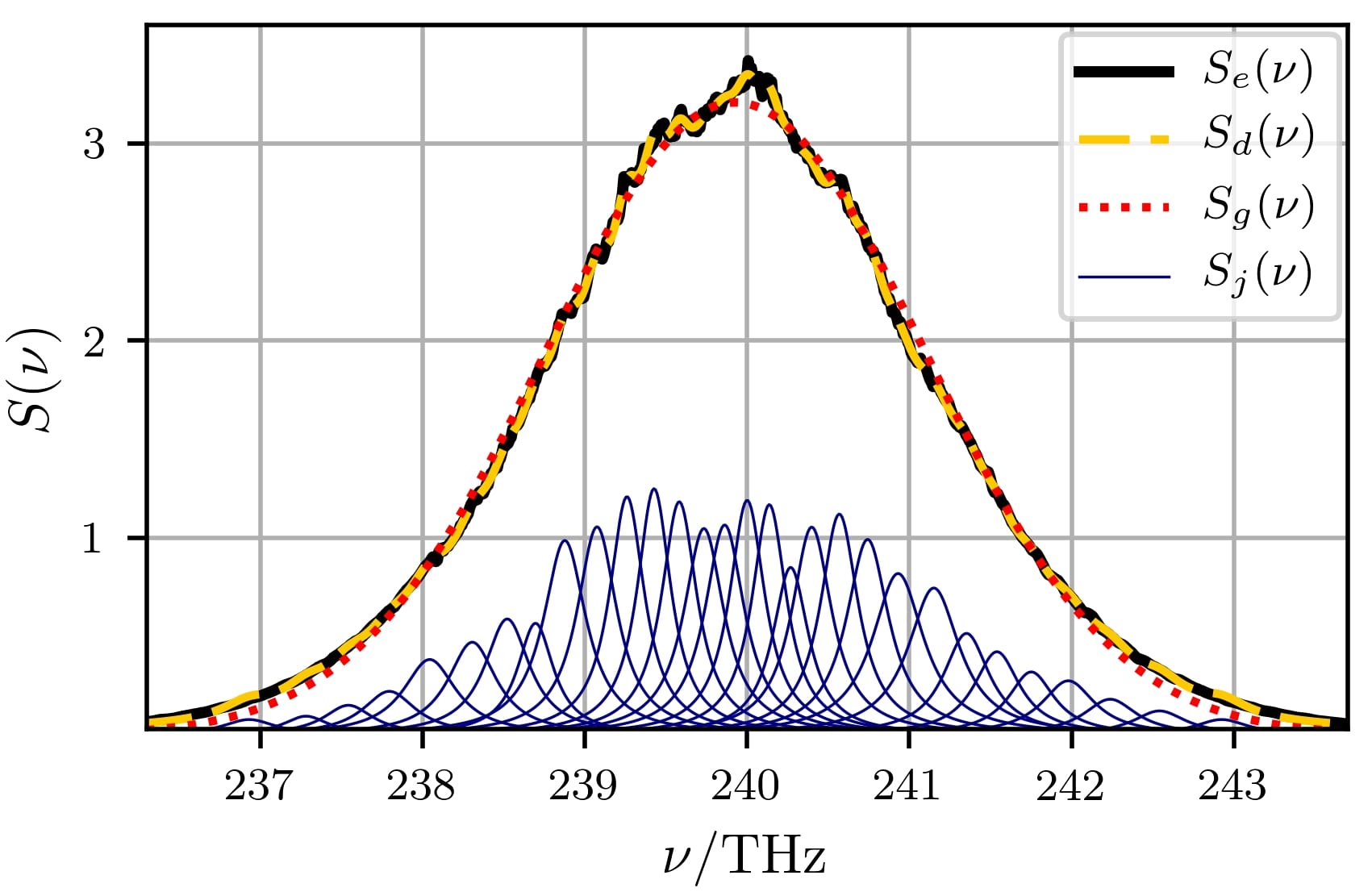}
	\caption{Experimental optical power spectrum $S_{\text{e}}(\nu)$ (black, solid) versus frequency $\nu$ \cite{Hartmann2017} with Gaussian fit $S_{\text{g}}(\nu)$ (red, dotted), stochastic emission $S_{\text{d}}(\nu)$ for $N=30$ oscillators according to eq. (\ref{eq:sumLorentizans}) (yellow, dashed) and spectra of individual oscillators $S_j(\nu)$ (solid, blue). The results of the Gaussian fit are a central frequency $\nu_0=239.9$\,THz and a width $\sigma=1.17$\,THz.}
	\label{fig:lorentzianModel}
\end{figure}

Superluminescent diodes are semiconductor-based light sources, which are characterized by spatially directed emission and spectral widths in the THz regime. The experiments with QDSLDs \cite{Hartmann2017} had an active medium consisting of inhomogeneously broadened InAs/InGaAs quantum dot layers. The optical power spectrum has a Gaussian shape (see Fig. \ref{fig:lorentzianModel}). The emission of the diode is modelled by $N=30$ individual oscillators using a least square fit (see eq. (\ref{eq:error})) to an experimental spectrum $S_{\text{e}}(\nu)$ \cite{Hartmann2017} to determine linewidths $\gamma_j$, mean intensities $I_j$ and central frequencies $\nu_j$ describing the emission. This is illustrated in Fig. \ref{fig:lorentzianModel}.

\section{QDSLD emission spectrum} \label{sec:spectrum}

The description of the QDSLD is used to simulate the output spectrum of such a diode. For this, the central frequencies $\nu_j$, linewidths $\gamma_j$ and mean intensities $I_j$ determined in Sec. \ref{model} are used to simulate the individual electric fields $\varepsilon_j(t)$ according to eq. (\ref{eq:simField}). The electric field produced by the diode $\varepsilon_{\text{d}}(t)$ results according to eq. (\ref{eq:diodeField}). Subsequently, the stationary first-order temporal correlation function $G_{\text{d}}^{(1)}(\tau)$ is calculated according to eq. (\ref{eq:simTempCor}) using $M=10^4$ realizations of the diode field $\varepsilon_{\text{d}}(t)$. The spectral power density of the emission $S_{\text{d}}(\nu)$ is determined using a Fourier transformation.

\begin{figure}[H]
	\centering
	\includegraphics[width=0.52\textwidth]{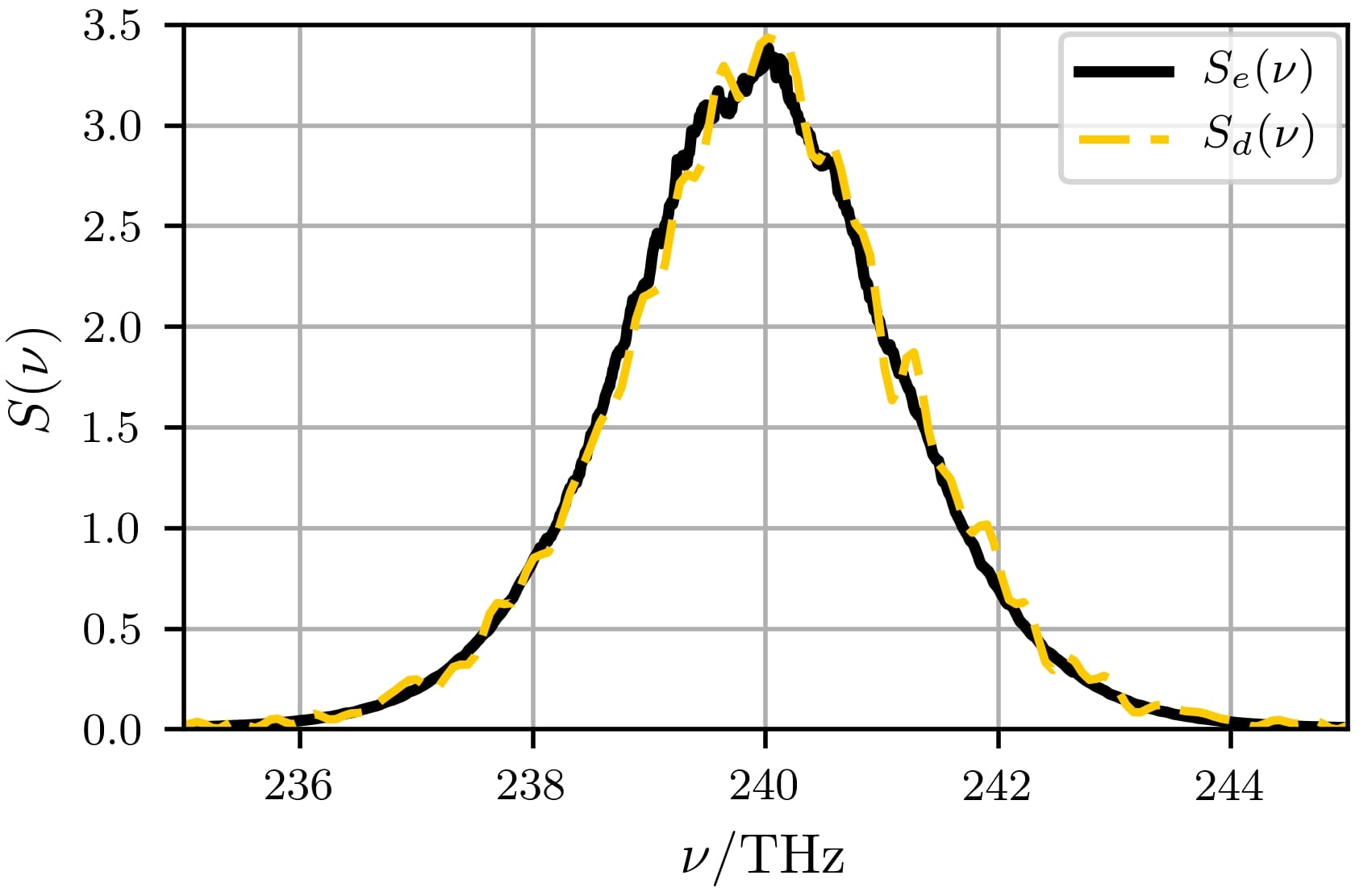}
	\caption{Experimental optical power spectrum $S_{\text{e}}(\nu)$ (black, solid) versus frequency $\nu$ \cite{Hartmann2017} with simulation results $S_{\text{d}}(\nu)$ (yellow, dashed) for $N=30$ oscillators resulting from a Fourier transformation of the stationary first-order temporal correlation function $G_{\text{d}}^{(1)}(\tau)$ calculated according to eq. (\ref{eq:simTempCor}) with $M=10^4$.}
	\label{fig:sumComplexGaussians}
\end{figure}

The result of the simulation (see Fig. \ref{fig:sumComplexGaussians}) shows good agreement with the experimental optical power spectrum \cite{Hartmann2017}. We define the width of the spectral power density \cite{Suessmann1997, Schleich2011}
\begin{align}
	b=\frac{1}{\int_{-\infty}^{\infty}\text{d}\nu\;S^2(\nu)}.
\end{align}
This yields $b_{\text{d}}=4.51\,\text{THz}$, implying a coherence time of $\tau_{\text{c,d}}=1/b_{\text{d}}=221.9\,\text{fs}$, which matches the experimental results of $b_{\text{e}}=4.29$\,THz and $\tau_{\text{c,e}}=233$\,fs very well. The developed formalism of modelling the emission of a quantum dot superluminescent diode as a superposition of individual oscillators is therefore suitable to describe the spectral power density of the diode.

\section{Second-order temporal correlation function} \label{sec:secOrdTemp}

In addition to the investigation of the optical power spectrum, the developed formalism can be used to investigate the classical photon statistics of the QDSLD emission. For this, $M$ realizations of the electric field $\varepsilon_{\text{d}}(t)$ determined in sec. \ref{sec:spectrum} are used to calculate the stationary normalized second-order temporal correlation function $g_{\text{d}}^{(2)}(\tau)$ of the emission (\ref{eq:secOrdTempAna}, \ref{eq:simTempCor}, \ref{eq:secOrdTempSim}).

\begin{figure}[H]
	\centering
	\includegraphics[width=0.52\textwidth]{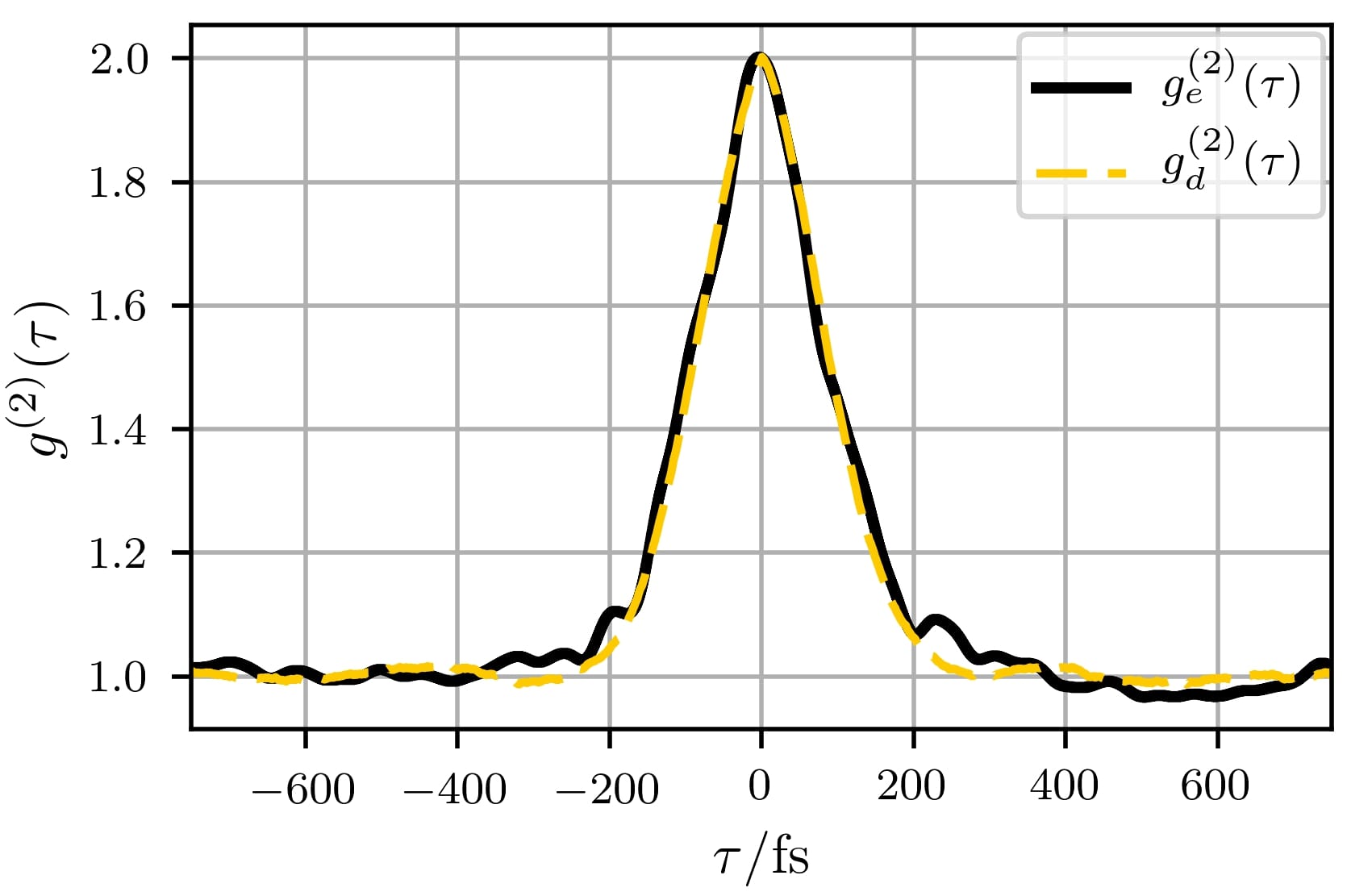}
	\caption{Experimental stationary normalized second-order temporal correlation function $g_{\text{e}}^{(2)}(\tau)$ (black, solid) versus time $\tau$ \cite{Hartmann2017} with simulation results $g_{\text{d}}^{(2)}(\tau)$ (yellow, dashed) for $N=30$ oscillators calculated according to eqs. (\ref{eq:simTempCor}, \ref{eq:secOrdTempSim}) with $M=10^4$.}
	\label{fig:secOrderTempCor}
\end{figure}

The result for the central frequencies $\nu_j$, linewidths $\gamma_j$ and mean intensities $I_j$ determined in section \ref{model} is illustrated in Fig. \ref{fig:secOrderTempCor}. There is good agreement between the simulation and the experimental data $g_{\text{e}}^{(2)}(\tau)$. The central degree of second-order temporal coherence $g_{\text{d}}^{(2)}(\tau=0)\simeq2$ indicates a Gaussian photon distribution, which was also shown experimentally. Therefore, the developed formalism is suited for classical photon statistical investigations of QDSLDs.

\section{Conclusion} \label{sec:conclusion}

In this article we study a stochastic model to describe experimental emission spectra. These are considered to result as a superposition of individual complex Ornstein-Uhlenbeck processes. The first- and second-order temporal correlation properties of these oscillators are investigated analytically and numerically. We can approximate Gaussian-, Lorentzian-, Voigt- and bandwidth limited spectra by determination of Lorentzian linewidths, carrier frequencies and amplitudes of the individual oscillators using least square fits.

The developed procedure is applied to the emission properties of quantum dot superluminescent diodes. Simulation parameters are extracted from a least square fit to an experimental spectrum \cite{Hartmann2017}. These are used to simulate the QDSLD emission and calculate first- and second-order temporal correlation properties. The determined spectral power density of the emission, resulting from a Fourier transformation of the stationary first-order temporal correlation function, shows good agreement with the experimental results regarding the shape of the spectral line, as well as spectral width and coherence time. Additionally, calculating the stationary normalized second-order temporal correlation function results in a a central degree of second-order temporal coherence $g^{(2)}(\tau=0)\simeq2$. This indicates a Gaussian photon distribution, which is in agreement with experiments and former theoretical investigations.

The stochastic description of QDSLD emission offers a straightforward perspective on the process of light generation inside QDSLDs, describing it as a superposition of individual classical oscillators. More data on the emission characteristics of the constituents of QDSLDs can lead to a better understanding and contribute to the design of new diodes. Furthermore, this approach can be used in the investigation of other phenomena shown by QDLSDs such as temperature dependent intensity noise reduction \cite{Blazek2011}.


\section*{Acknowledgement}
\addcontentsline{toc}{section}{Acknowledgement}

We thank S\'{e}bastien Blumenstein for the provision of experimental data and \mbox{Prof. Wolfgang} Els\"{a}\ss er for stimulating discussions.

\appendix
\section{Convergence of stochastic simulations} \label{sec:numSimSDE}

Consider the Ito stochastic differential equation \cite{Gardiner2009}
\begin{align}
	\text{d}x(t)=a(x(t))\,\text{d}t+b(x(t))\,\text{d}W(t),
\end{align}
with the drift term $a(x)$ and the diffusion term $b(x)$. Identifying $x(t_0)=x_0$, the formal solution of this equation is given by integration:
\begin{align}
	x(t)=x_0+\int_{t_0}^{t}\text{d}t'\,a(x(t'))+\int_{t_0}^{t}\text{d}W(t')\,b(x(t'))\label{eq:formalSolIto}
\end{align}

The goal of time discrete maps $x(t_{i+1})=F(x_i)$ of stochastic differential equations is the approximation of a solution $x(t)$ up to a order of convergence $\gamma$. Such a scheme is said to converge strongly with order $\gamma>0$, if for the final time instant $T$ and $N=T/\Delta$ there is a finite $\epsilon$ and $\Delta_0>0$ such that \cite{Kloeden1999}
\begin{align}
	\left\langle \vert x(T)-x(t_{N})\vert\right\rangle\leq\epsilon\Delta^{\gamma}
\end{align}
for any time discretization $0<\Delta<\Delta_0$. A strong Taylor scheme of order $\gamma$ can be constructed by considering the Ito-Taylor expansion, which is obtained by continuously applying the integral form of Ito's formula \cite{Gardiner2009}
\begin{align}
	f(x(t))=f(x_0)+\int_{t_0}^{t}\text{d}t'\,L^0f(x(t'))+\int_{t_0}^t\text{d}W(t')\;L^1f(x(t')),
\end{align}
with $L^0=a(x(t'))\partial_x+(1/2)b^2(x(t'))\partial_x^2$ and $L^1=b(x(t'))\partial_x$, to nonconstant terms inside the integrals of the formal solution (\ref{eq:formalSolIto}). A criterium \cite{Kloeden1999} for the terms of the Ito-Taylor expansion required for the associated strong Taylor scheme to achieve a desired order of strong convergence $\gamma$ states, that a simulation scheme strongly converges to the order of an integer $\gamma$ if it includes all combinations of integrals up to this order, with time differentials $\text{d}t$ being of order $1$ and Wiener noise increments $\text{d}W(t)$ being of order $1/2$. Simulation schemes of half-integer order $\gamma$ additionally require the inclusion of the pure time integral of order $\gamma+1/2$.

A strong convergence scheme of order $1/2$ is the Euler scheme
\begin{align}
	x(t)=x_0+a(x_0)\int_{t_0}^{t}\text{d}t'+b(x_0)\int_{t_0}^{t}\text{d}W(t')+\mathcal{R}.\label{eq:eulerInt}
\end{align}
Discretizing the time steps, discrete map can be developed which yields
\begin{align}
	x(t_{i+1})=x(t_i)+a(x(t_i))\,\Delta t+b(x(t_i))\,\Delta W,
\end{align}
where $\Delta W$ is a Gaussian random process with $\langle\Delta W\rangle=0$ and $\langle\Delta W^2\rangle=\Delta t$. To expand the Euler scheme to order $1.0$ of strong convergence, the double stochastic integral appearing in the remainder $\mathcal{R}$ in eq. (\ref{eq:eulerInt}) has to be included, yielding
\begin{align}
	x(t)=x_0+a(x_0)\int_{t_0}^{t}\text{d}t'+b(x_0)\int_{t_0}^{t}\text{d}W(t')+L^1b(x_0)\int_{t_0}^{t}\int_{t_0}^{t'}\text{d}W(t')\text{d}W(t'')+\mathcal{R}.
\end{align}
This is called the Milstein scheme \cite{Milstein1975}. The double stochastic integral can be calculated \cite{Gardiner2009}
\begin{align}
	\int_{t_0}^{t}\int_{t_0}^{t'}\text{d}W(t')\text{d}W(t'')=\frac{1}{2}\left[\left(W(t)-W(t_0)\right)^2-\left(t-t_0\right)\right],
\end{align}
which leads to the iteration rule for the Milstein method
\begin{align}
	x(t_{i+1})=&x(t_i)+\left[a(x(t_i))-\frac{1}{2}b(x(t_i))\partial_x b(x(t_i))\right]\Delta t\nonumber\\&+b(x(t_i))\,\Delta W+\frac{1}{2}b(x(t_i))\partial_x b(x(t_i))\,\Delta W^2.
\end{align}
With increasing order of convergence $\gamma$, the simulation schemes become more complex and include an increasing number of stochastic increments.


\def\cprime{$'$} \def\cprime{$'$}


\end{document}